# Finite size effects of hysteretic dynamics in multi-layer graphene on ferroelectric

Anna N. Morozovska[1], Anastasiya S. Pusenkova[2], Oleksandr V. Varenyk[1],

Sergei V. Kalinin[3], Eugene A. Eliseev[4], and Maxim V. Strikha[5*]

[1]*Institute of Physics, National Academy of Sciences of Ukraine, Kyiv, Ukraine*

[2]*Taras Shevchenko Kyiv National University, Physics Faculty, Kyiv, Ukraine*

[3] *The Center for Nanophase Materials Sciences and Materials Science and Technology Division, Oak Ridge National Laboratory, Oak Ridge, TN 37831*

[4]*Institute for Problems of Material Sciences, National Academy of Sciences of Ukraine, Kyiv, Ukraine*

[5] *V.E. Lashkarev Institute of Semiconductor Physics, National Academy of Sciences of Ukraine, Kyiv, Ukraine*

## Abstract

The origin and influence of finite size effects on the nonlinear dynamics of space charge stored by multi-layer graphene on ferroelectric and the resistivity of graphene's channel were analyzed using self-consistent continuum media approach. Revealed size effects peculiarities are governed by the relations between the thicknesses of multi-layer graphene, ferroelectric film and dielectric layer between them. Appearance of charge and electro-resistance hysteresis loops and their versatility steam from the interplay of polarization reversal dynamics in alternating electric field and its incomplete screening, which features are mostly determined by the dielectric layer thickness. We further investigate the effects of spatially non-uniform ferroelectric domain structures on graphene layer conductivity and predict its dramatic increase under the transition from multi- to single- domain state in ferroelectric. The intriguing effects can open new possibilities for graphene-based sensors and explore the physical mechanisms underlying the operation of graphene field effect transistor with ferroelectric gating.

---

[*] Corresponding author e-mail: maksym_strikha@hotmail.com

# 1. Introduction

Since graphene fabrication in early 2004, the new rapidly progressing inter-disciple, graphene physics, was born; it is closely related with nanophysics, borders with condensed matter and surface physics, physical chemistry and engineering [1, 2, 3, 4]. From the beginning, graphene sheets attracted most of experimental and theoretical interest, without the regard for the physical nature of the substrate they were placed on etc. Then physical understanding occurs that graphene can have fundamental and applied interest together with other functional components, such as its interaction with different substrate, gates, contacts, thermostat, and other important factors that determine the electro-transport peculiarities.

The study of graphene on the substrates with high permittivity appeared of great interest, because the substrates permit to reach higher carrier concentrations for the same gate voltages [4, 5, 6]. In particular, usage of ferroelectrics (which permittivity is typically at least one order of magnitude higher than that for linear dielectrics) instead of traditional high k substrates allows controlling space charge density distribution in single and multi-layered graphene by changing the spontaneous polarization direction, value and domain structure properties of ferroelectric [7, 8, 9, 10, 11, 12, 13, 14]. Charge carriers in graphene screen the depolarization electric field, caused by spontaneous polarization at the ferroelectric surface, and the sign of the screening carriers depends on the spontaneous polarization direction [15, 16]. Allowing for the screening mechanism ferroelectric substrate makes graphene highly sensitive to electric field, elastic strain and temperature. The use of ferroelectric substrates like organic relaxor PVDF-TrFE, which behave at low voltage as paraelectrics with extremely high dielectric permittivity, can be beneficial for the construction of low-voltage near- and mid-IR range modulators [17]. Moreover, the existence of hysteresis in ferroelectric polarization dependence on the applied field make them promising for the construction of memory units [7]. The first study focused on graphene-on-ferroelectric device had appeared in 2009 [7], and the number of such works permanently increases, and several reviews were already published (see e.g. [18, 19, 20]).

Despite recent experiments on ferroelectric gating [7, 10, 12, 13] revealed a novel functionality, nonvolatility, in graphene field-effect transistors (GFeFETs), a comprehensive understanding in the nonlinear, hysteretic ferroelectric gating and an effective way to control it are still incomplete [9]. Zheng et al. characterized the hysteretic ferroelectric gating and demonstrated symmetrical bit writing in GFeFETs with electro-resistance change over 500% and reproducible no-volatile switching and have demonstrated. However the self-consistent influence of depolarization field on graphene layer, separated from ferroelectric substrate by an air or dielectric gap, as well as finite size effects impact have not been studied earlier.

This motivate us evolve in this work a continuum media theory of the size effects influence on the storied charge and electro-resistance nonlinear hysteretic dynamics in multi-layer graphene on ferroelectric. Also we aimed to establish the impact of buffer dielectric layer between graphene and ferroelectric surface with domain structure. We consider multi-layer graphene layer of finite thickness and dynamic hysteretic effects, in comparison with our previous studies [16], where we study the thermodynamics of charge density in semi-infinite multi-layer graphene.

## 2. Problem statement

Geometry of the considered GFeFET with two gates (similar geometry of the problem was used in [7, 9]) is shown in the **Figure 1.** Multi-layered graphene (MLG) of thickness $d$ has background permittivity $\varepsilon_G$. Ultra-thin dielectric layer has thickness $h$ and dielectric permittivity $\varepsilon_d$ (this imposes limits on $h$, which should be at least of several lattice constants order, because for smaller $h$ permittivity $\varepsilon_d$ trends to zero). Ferroelectric film with 180-degree domain structure and spontaneous polarization vector $\mathbf{P}_S = (0,0,P_3)$ has thickness $l$. The period of 180-degree domain structure is $a$. There is a top gate electrode under $ac$ voltage $V_{TG}$; below there is a bottom gate electrode under the $dc$ voltage $V_{BG}$; they dopes MLG with electrons due to capacitor effect, the voltage on the gate determines the Fermi energy level $E_F$ position. The driving field $E_S$ can be applied to the MLG channel.

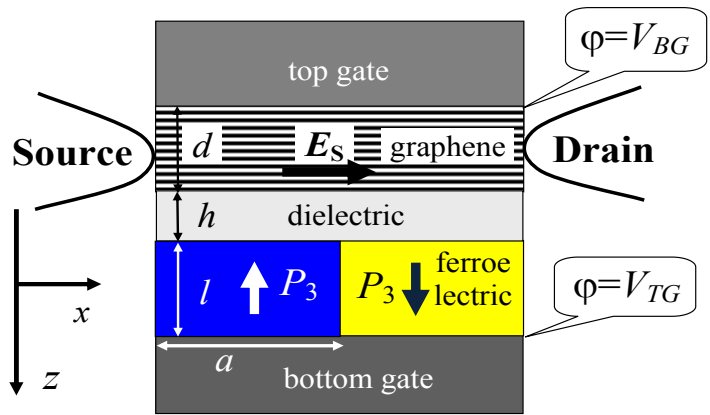

**Figure 1.** Geometry of the FET heterostructure, top gate/MLG/dielectric layer/ferroelectric /bottom gate.

The system of electrostatic equations for the given problem is:

$$\left(\frac{\partial^2}{\partial z^2}+\frac{\partial^2}{\partial x^2}+\frac{\partial^2}{\partial y^2}\right)\varphi_G-\frac{\varphi_G}{R_t^2}=0, \quad \text{for} \quad -d<z<0, \quad \text{(graphene)} \tag{1a}$$

$$\left(\frac{\partial^2}{\partial z^2}+\frac{\partial^2}{\partial x^2}+\frac{\partial^2}{\partial y^2}\right)\varphi_d=0, \quad \text{for} \quad 0<z<h, \quad \text{(dielectric layer)} \tag{1b}$$

$$\left(\varepsilon_{33}^f\frac{\partial^2}{\partial z^2}+\varepsilon_{11}^f\left(\frac{\partial^2}{\partial x^2}+\frac{\partial^2}{\partial y^2}\right)\right)\varphi_f=0, \quad \text{for} \quad h<z<L. \quad \text{(ferroelectric film)} \tag{1c}$$

Thickness $L=l+h$, $\varepsilon_{ij}^f$ are the linear dielectric permittivity tensor components, $R_t$ is a screening radius [21], which value can be estimated using Debye or Tomas-Fermi approximations depending on the heterostructure electronic properties and finite size effects.

Equations (1) are supplemented by the boundary conditions of electric potential φ and electric displacement **D** normal components continuity [15] at the interfaces $z=-d$, $z=0$, $z=h$ and $z=L$. For the case when electric current along graphene channel is absent ($E_S=0$), but nonzero top and bottom gate voltages are applied, they acquire the following form:

$$\varphi_G(x,y,z=-d)=V_{TG}, \tag{2a}$$

$$\varphi_G(x,y,0)=\varphi_d(x,y,0), \tag{2b}$$

$$D_3^G(x,y,0)-D_3^d(x,y,0)=0 \;\Rightarrow\; \varepsilon_0\left(-\varepsilon_G\frac{\partial\varphi_G(x,y,0)}{\partial z}+\varepsilon_d\frac{\partial\varphi_d(x,y,0)}{\partial z}\right)=0, \tag{2c}$$

$$\varphi_d(x,y,h)=\varphi_f(x,y,h), \tag{2d}$$

$$D_3^f-D_3^d=0 \;\Rightarrow\; -\varepsilon_0\varepsilon_{33}^f\frac{\partial\varphi_f(x,y,h)}{\partial z}+P_3(x,y)+\varepsilon_0\varepsilon_d\frac{\partial\varphi_d(x,y,h)}{\partial z}=0, \tag{2e}$$

$$\varphi_f(x,y,z=L)=V_{BG}. \tag{2f}$$

Here universal dielectric constant $\varepsilon_0=8.85\times10^{-12}$ F/m. Displacement is related with ferroelectric polarization as $D_3^f=\varepsilon_0\varepsilon_{33}^f E_3^f+P_3$. In dielectric $D_3^d=\varepsilon_0\varepsilon^d E_3^d$ and in graphene $D_3^G=\varepsilon_0\varepsilon^G E_3^G$. In ferroelectric the polarization should be found from the time-dependent Landau-Ginzburg-Devonshire (LGD) equation of state

$$\Gamma\frac{\partial}{\partial t}P_3+\alpha P_3+\beta P_3^3+\gamma P_3^5-g\left(\frac{\partial^2}{\partial z^2}+\frac{\partial^2}{\partial x^2}+\frac{\partial^2}{\partial y^2}\right)P_3=E_3^f, \tag{3}$$

supplemented by e.g. natural boundary conditions, $\left(\partial P_3/\partial z\right)\big|_{z=h,L}=0$.

## 3. Solution for electric potential in a single-domain state of ferroelectric

The contribution of the top and bottom gate potentials $V_{TG}$ and $V_{BG}$ and polarization $P_3$ into the solution of the boundary problem (1)-(2) is given by expressions:

$$\varphi_G(z) = \frac{\begin{pmatrix} V_{TG}\left(e^{(d+z)/R_t}\varepsilon_d\varepsilon_G l\left(e^{-2z/R_t}+1\right)+\varepsilon_{33}^f\left(\varepsilon_G h\left(e^{-2z/R_t}+1\right)+\varepsilon_d R_t\left(e^{-2z/R_t}-1\right)\right)\right) \\ +\varepsilon_d R_t\left(\varepsilon_{33}^f V_{BG}-(P_3 l/\varepsilon_0)\right)\left(e^{(2d+z)/R_t}-e^{-z/R_t}\right) \end{pmatrix}}{\varepsilon_d\varepsilon_G l\left(e^{2d/R_t}+1\right)+\varepsilon_{33}^f\left(\varepsilon_G h\left(e^{2d/R_t}+1\right)+\varepsilon_d R_t\left(e^{2d/R_t}-1\right)\right)}, \quad (4a)$$

$$\varphi_d(z) = \frac{2V_{TG}e^{d/R_t}\left(\varepsilon_d l+\varepsilon_{33}^f(h-z)\right)\varepsilon_G+\left(\varepsilon_{33}^f V_{BG}-(P_3 l/\varepsilon_0)\right)\left(\varepsilon_d R_t\left(e^{2d/R_t}-1\right)+\varepsilon_G z\left(e^{2d/R_t}+1\right)\right)}{\varepsilon_d\varepsilon_G l\left(e^{2d/R_t}+1\right)+\varepsilon_{33}^f\left(\varepsilon_G h\left(e^{2d/R_t}+1\right)+\varepsilon_d R_t\left(e^{2d/R_t}-1\right)\right)}, \quad (4b)$$

$$\varphi_f(z) = V_{BG}+(h+l-z)\frac{\varepsilon_d\varepsilon_G\left(2V_{TG}e^{d/R_t}-V_{BG}\left(e^{2d/R_t}+1\right)\right)-\left(\varepsilon_G h\left(e^{2d/R_t}+1\right)+\varepsilon_d R_t\left(e^{2d/R_t}-1\right)\right)(P_3/\varepsilon_0)}{\varepsilon_d\varepsilon_G l\left(e^{2d/R_t}+1\right)+\varepsilon_{33}^f\left(\varepsilon_G h\left(e^{2d/R_t}+1\right)+\varepsilon_d R_t\left(e^{2d/R_t}-1\right)\right)}$$

(4c)

where the subscript $j = G, d, f$ means graphene, dielectric and ferroelectric correspondingly. Using Eq.(4c) the electric field in ferroelectric can be derived as:

$$E_3^f = \frac{\varepsilon_d\varepsilon_G\left(2V_{TG}e^{d/R_t}-V_{BG}\left(e^{2d/R_t}+1\right)\right)-\left(\varepsilon_G h\left(e^{2d/R_t}+1\right)+\varepsilon_d R_t\left(e^{2d/R_t}-1\right)\right)(P_3/\varepsilon_0)}{\varepsilon_d\varepsilon_G l\left(e^{2d/R_t}+1\right)+\varepsilon_{33}^f\left(\varepsilon_G h\left(e^{2d/R_t}+1\right)+\varepsilon_d R_t\left(e^{2d/R_t}-1\right)\right)} \quad (5)$$

The field is constant with z coordinate, however it depends not only on the screening radius $R_t$ but also on polarization $P_3$, which in its turn depends on $R_t$ via electric field inside ferroelectric layer.

Since the electric field is constant in ferroelectric for the single-domain case, the solution of the TLGD equation of state (3) becomes consistent with natural boundary conditions and should be obtained self-consistently from the differential (or algebraic in the stationary case) equation

$$\Gamma\frac{\partial}{\partial t}P_3+\alpha P_3+\beta P_3^3+\gamma P_3^5 = E_3^f \quad (6)$$

Using a screening radius approximation Eq.(2a) (see also [2]) the space charge density $\rho_G(z)$ in the multi-layered graphene is given by the expressions:

$$\rho_G(z) = -\varepsilon_0\varepsilon_G\frac{\varphi_G(z)}{R_t^2} \quad (7)$$

The total charge $\sigma_G$ is the sum of the integrated over graphene thickness space charge density $\rho_G(z)$ and the surface charge determined by the boundary condition at the graphene-top gate interface $z=-d$ as $D_3^G(-d)=\sigma_S$. In accordance with the principle of the whole system electroneutrality the total charge stored in graphene is opposite in sign to the electric displacement at the bottom electrode, $D_3^f\big|_{z=L} = \left(\varepsilon_0\varepsilon_{33}^f E_3^f+P_3\right)\big|_{z=L}$, that in allowance for Eq.(5) gives:

$$\sigma_G = \varepsilon_0 \varepsilon_d \varepsilon_G \frac{2\varepsilon_{33}^f V_{TG} e^{d/R_t} - \left(\varepsilon_{33}^f V_{BG} - (P_3 l/\varepsilon_0)\right)\left(e^{2d/R_t} + 1\right)}{\varepsilon_d \varepsilon_G l\left(e^{2d/R_t} + 1\right) + \varepsilon_{33}^f \left(\varepsilon_G h\left(e^{2d/R_t} + 1\right) + \varepsilon_d R_t \left(e^{2d/R_t} - 1\right)\right)} \quad (8)$$

Note that the Eqs.(5)-(6) are coupled, and their self-consistent solution for polarization should be substituted into Eq.(8) to calculate the total charge of graphene. Then the effective surface density of carriers in graphene could be found as $n = -(\sigma_G/e)$ [3], $e = 1.6 \times 10^{-19}$ C is the electron charge. Finally the graphene "effective" electro-resistance $R$ can be estimated (see e.g. [9, 13]) as:

$$R = \frac{R_0}{\sqrt{1 + n^2/n_{res}^2}} \quad (9)$$

Here $R_0 = \frac{L_{SD}}{We\mu_{Hall} n_{res}}$, $L_{SD}$ is graphene's channel length from source to drain, $W$ is it's width, and $n_{res}$ is a residual carriers density in graphene.

### 4. Nonlinear hysteretic effect on the graphene conductivity and electro-resistance

In this section we analyze the dynamics of the space charge storied in graphene when ferroelectric film is in a single-domain case. In accordance with a relevant experiment [7], bottom gate voltage is constant and the top gate voltage varies with frequency ω, i.e. $V_{TG} = V_{TG} \sin(\omega t)$. Dimensionless frequency is introduced as $w = -\omega\Gamma/\alpha$. Full set of parameters are listed in the **Table 1**.

**Table 1. Range of parameters used in calculations**

| Parameters | Value |
|---|---|
| screening radius $R_t$ | (0.3 - 3) nm |
| graphite permittivity $\varepsilon_G$ | 15 |
| graphene thickness $d$ | (3 – 30) nm |
| residual concentration $n_{res}$ | (vary from $10^{17}$ m$^{-2}$ to $10^{15}$ m$^{-2}$, as the latter value was used in [9]) |
| dielectric permittivity $\varepsilon_d$ | 1(air); 5-7 (background constant), 12.53 (sapphire Al$_2$O$_3$) |
| dielectric layer thickness $h$ | (0 – 50) nm |
| dielectric anisotropy of ferroelectric $\gamma$ | 0.58 (LiNbO$_3$); 3.87 (Rochelle salt) |
| ferroelectric permittivity $\varepsilon_{33}^f$ | 29 (LiNbO$_3$); 300 (Rochelle salt) |
| ferroelectric polarization $P_S$ | 0.75 C/m$^2$ (LiNbO$_3$); 0.002 C/m$^2$ (Rochelle salt or relaxor) |
| LGD parameters for LiNbO$_3$ | $\alpha = -1.95 \times 10^9$ m/F, $\beta = 3.61 \times 10^9$ m$^5$/(C$^2$F), $\gamma = 0$ |
| ferroelectric thickness $l$ | (10 – 30) nm |
| period of domain structure $a$ | (50 – 500) nm |
| coercive field range | $(2 – 7) \times 10^8$ V/m (LiNbO$_3$); 30 kV/cm (Rochelle salt) |

**Figures 2-5** show ferroelectric polarization $P_3$, effective density of total charge stored in graphene $n$ and its electro-resistance $R$ as a function of the top gate voltages $V_{TG}$ at several fixed values of $V_{BG}$, different ferroelectric substrate thickness $l$, dielectric layer thickness $h$, graphene thickness $d$,

and screening radius $R_t$. Left column of the Figs.2-4 (plots a, c, e) are calculated for the lower frequency $w = 0.03$ and $V_{BG} = +5$ V; the right column (plots b, d, f) corresponds to the lower frequency $w = 0.3$ and $V_{BG} = +5$ V. Nonlinear and hysteresis effects originate, when electric field in ferroelectric $E_3^f$ is enough to reverse or significantly change its polarization $P_3$ in agreement with Eq.(6). Polarization hysteresis immediately causes the hysteretic response of the total charge stored in multi-layer graphene and its resistance. At that the heterostructure response can be asymmetric, since initial polarization direction exists. Generally these results correspond to the mechanism of the "direct" hysteresis in graphene's channel resistivity, caused by the re-polarization of the substrate dipoles (see e.g. [22]).

General feature is that the strong "inflation" of polarization and graphene charge hysteresis loops under the frequency increase in one order in magnitude. The term "inflation" means here essential increase of the coercive voltage, which value in fact defines the memory window, remnant charge and polarization, as well as the strong rounding of the loops shape. The electro-resistance loops behaviour with frequency increase is nontrivial, but rounding effect is pronounced. As anticipated the bottom gate voltage $V_{BG}$ causes the constant horizontal shift (offset) of the hysteresis loops along the top gate axes $V_{TG}$. Positive $V_{BG}$ leads to the left offset of the loops, negative $V_{BG}$ leads to right one. The higher is the $V_{BG}$ value the stronger is the offset. The graphene charge loop shape, remnant charge and coercive voltage values strongly correlate with the polarization loop, but the parameters are different, since the charge is proportional to $P_3$ and $V_{BG}$ in accordance with Eq.(8), namely $\sigma_G = C_1 V_{TG} + C_2 V_{BG} + C_3 P_3$. Moreover, non-trivial vertical and horizontal asymmetry of all loops originates from $V_{BG}$ impact; it is absent at $V_{BG}=0$. The "depth" of the graphene electro-resistance response modulation is defined by the residual carrier density in graphene, $n_{res}$, and readily reaches one order of magnitude at chosen parameters. It strongly increases with $n_{res}$ decrease in agreement with Eq.(9).

**Figures 2-4** analyses allow us to receive the simplest notion about the influence of finite size effects on the ferroelectric polarization, storied charge and graphene effective electro-resistance. To receive the information let us analyze the dependence of $P_3$, $\sigma_G$ and $R$ voltage response on ferroelectric substrate thickness $l$, dielectric layer thickness $h$ and graphene thickness $d$.

Coercive voltage and remnant polarization values for $P_3$ and $\sigma_G$ loops noticeably decrease with $l$ decrease from 30 nm to 15 nm [curves 1-4 in **Figs.2**]. Coercive voltage and remnant polarization values for $P_3$ and $\sigma_G$ loops strongly decrease with dielectric layer thickness $h$ increase from 0 (intimate contact) to 10 nm [curves 1-4 in **Figs 3**]. The loops loses the squire-like and acquires slim-like shape with $l$ decrease or $h$ increase, that is characteristic feature for when approaching the transition to paraelectric phase. Maxima on graphene resistance loops approach each other and tend to merge under $l$ decrease or $h$ increase. This decrease and loop shape degradation occurs from the depolarization field, which impact increases with $l$ decrease (or $h$ increase) leading to the film ferroelectric properties degradation up to their disappearance at $l < l_{cr}$, where the critical thickness $l_{cr}$ depends on $h$, $d$ and $R_t$, dielectric permittivities as well as on the ferroelectric material parameter α. For chosen parameters $l_{cr}$ is about 10 nm. Expression for depolarization field follows from Eq.(5) at $V_{TG}=V_{BG}=0$, namely

$$E_3^d = \frac{-\left(\varepsilon_G h\left(e^{2d/R_t}+1\right)+\varepsilon_d R_t\left(e^{2d/R_t}-1\right)\right)(P_3/\varepsilon_0)}{\varepsilon_d \varepsilon_G l\left(e^{2d/R_t}+1\right)+\varepsilon_{33}^f\left(\varepsilon_G h\left(e^{2d/R_t}+1\right)+\varepsilon_d R_t\left(e^{2d/R_t}-1\right)\right)}. \quad (10)$$

As one can see from the expression, $l$ decrease leads to the monotonic increase of depolarization field. The increase of $h$ firstly leads to the strong increase of the depolarization field and then to its value saturation, since $E_3^d \propto \frac{C_1 h + C_2}{C_3 h + C_4}$. The ferroelectricity degradation or disappearance in the film leads to the simultaneous degradation or disappearance of the graphene charge and resistance hysteresis. So, the optimization between the aim to reach the system miniaturization due to ferroelectric film thickness decrease and to conserve its high polar-active properties is required. Also the field does not depend on MLG thickness $d$ separately, but depends on the ratio $2d/R_t$.

Polarization, storied charge and graphene resistance loops shape, asymmetry and other parameters change in a non-trivial way under graphene thickness $d$ two times increase from 3 nm to 6 nm (from approximately 10 to 20 graphene layers, [curves 1-4 in **Figs 4**]). With $d$ increase polarization and charge loops becomes more slim, strongly tilted and shifted at low frequencies and wide dielectric gap $h$ =10 nm. The coercive voltages decrease with $d$ increase, but not so strongly as in the previous cases. At higher frequencies polarization and total charge loops acquire quasi-elliptic shape with noticeable vertical and horizontal asymmetry originated from $V_{BG}$ application. Remarkably, the loops asymmetry increases and resistance loops double shape transforms into a single one with $d$ increase. The strong influence of graphene thickness $d$ on the heterostructure voltage response partially comes from the complex exponential dependence of the depolarization field on the parameter.

The resistance loop shape shown in **Figs.4f** are versatile and very similar to the experimental ones obtained by Zheng et al (Fig.2 in [9]), while the amplitude of ac voltage applied to the top gate ($V_{TG0}$=30 V) was taken much higher than the bottom gate voltage ($V_{BG}$ =± 5 V). In fact we used high $V_{TG0}$ in order to reverse the spontaneous polarization of ferroelectric, which hysteresis cases the memory effect in graphene.

Note, that analytical results illustrated by **Figs.2-4** are complementary to the experiment and modeling performed by Zheng et al, because we analysed the influence of the heterostructure finite sizes on graphene sheet conductivity and resistance.

**Figures 5** illustrate changes of polarization, storied charge and resistance hysteresis loops, which appear for different screening radiuses $R_t$. Coercive voltage value for $P_3$ increases with increasing $R_t$, while remnant polarization decreases. Polarization loop quickly disappears when multi-layer graphene thickness $d$ becomes higher than $2R_t$, because electric field in MLG attenuates [black curves "1" in **Figs. 5a,b**]. Total charge density value accumulated by graphene and charge loop width become smaller when the screening radius becomes bigger [curves 1-4 in **Fig 5c,d**]. Furthermore, the distance between two resistance peaks become smaller ("memory window" decreases) and curves become smoother with $R_t$ increase [curves 1-4 in **Fig 5e,f**]. For high frequency resistance loops double shape transforms into a single one with $R_t$ increases.

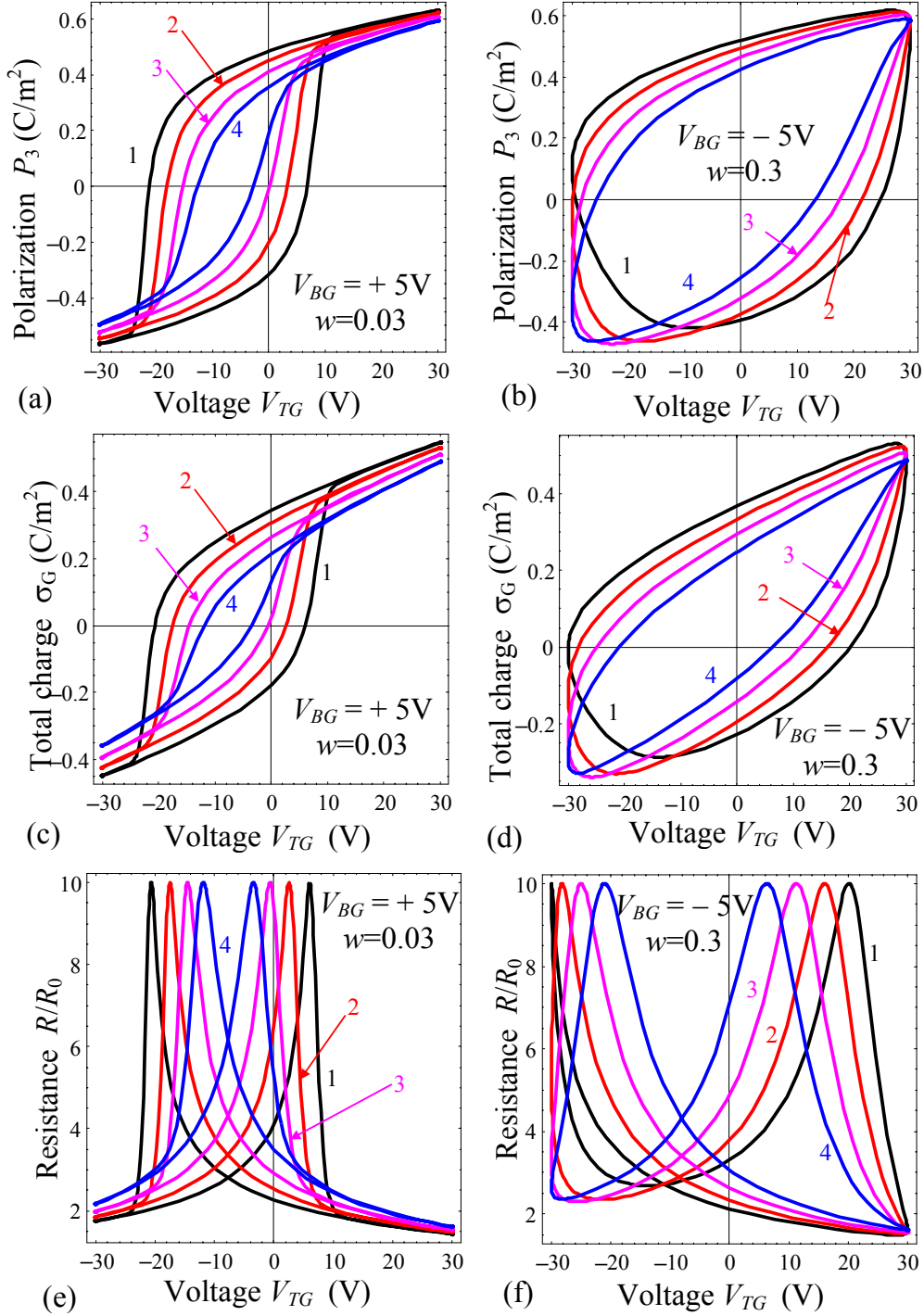

**Figure 2. Influence of ferroelectric film thickness.** Hysteresis loops of LiNbO$_3$ ferroelectric polarization **(a, b)**, density of the total charge stored in multi-layer graphene **(c, d)** and its electro-resistance **(e, f)** calculated for different thickness of ferroelectric film $l$=30, 25, 20, 15 nm (curves 1, 2, 3, 4). The frequency w=0.03 and back gate voltage $V_{BG}$= + 5 V for plots **(a)**, **(c)** and **(e)** (left column); w=0.3 and back gate voltage $V_{BG}$= − 5 V for plots **(b)**, **(d)** and **(f)** (right column). Sapphire dielectric thickness $h$=10 nm, multi-layer graphene thickness $d$=3 nm, screening radius $R_t$=3 nm, $n_{res}$ = 5×10$^{17}$ m$^{-2}$.

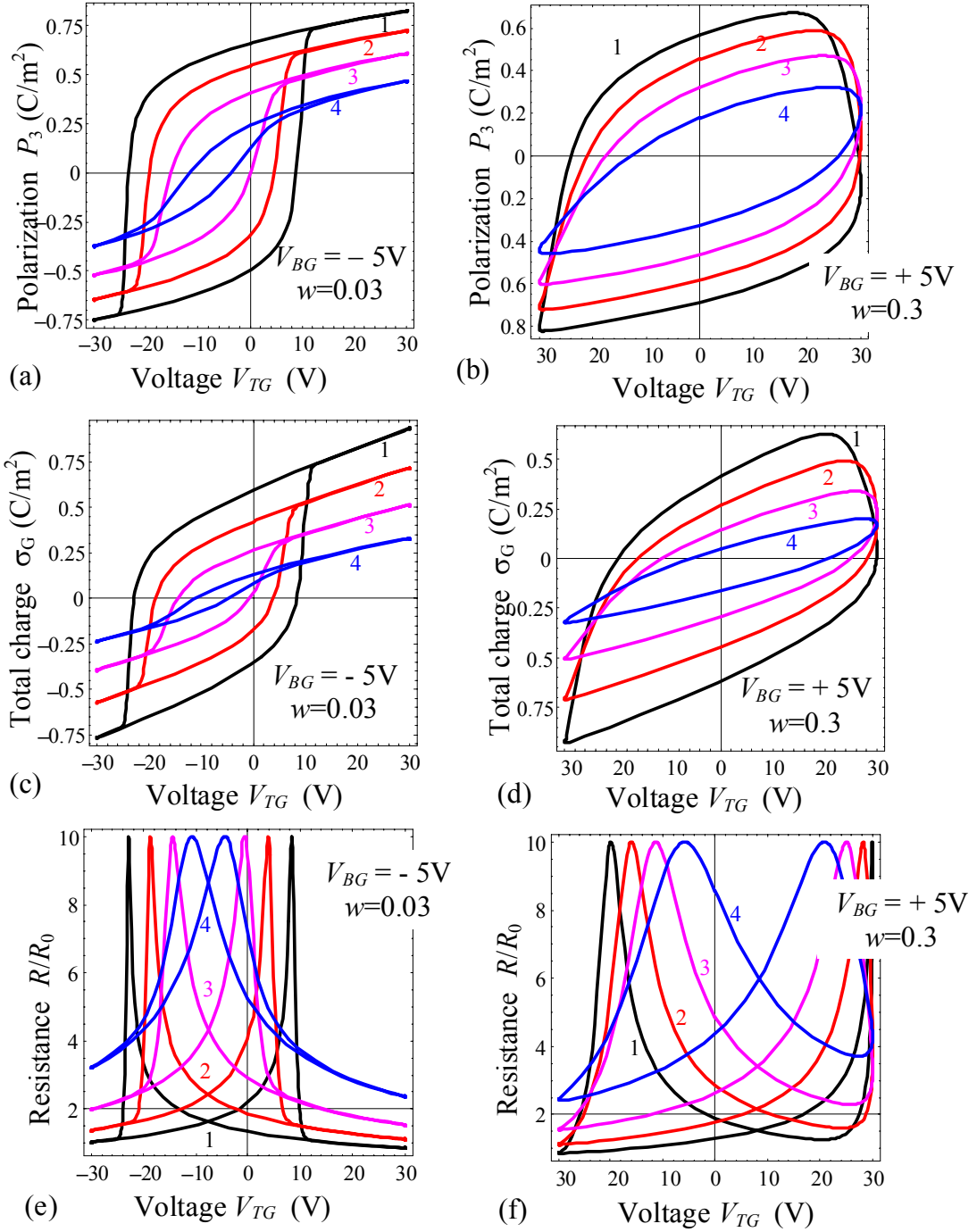

**Figure 3. Influence of dielectric layer thickness.** Hysteresis loops of LiNbO$_3$ ferroelectric polarization **(a, b)**, density of the total charge stored in multi-layer graphene **(c, d)** and its electro-resistance **(e, f)** calculated for different thickness of sapphire dielectric layer $h$=0, 2, 5, 10 nm (curves 1, 2, 3, 4). The frequency w=0.03 and back gate voltage $V_{BG}$= - 5 V for plots **(a)**, **(c)** and **(e)** (left column); w=0.3 and back gate voltage $V_{BG}$= + 5 V for plots **(b)**, **(d)** and **(f)** (right column). Multi-layer graphene thickness $d$=3 nm, screening radius $R_t$=3 nm, $n_{res}$ = 5×10$^{17}$ m$^{-2}$, ferroelectric film thickness $l$=20 nm.

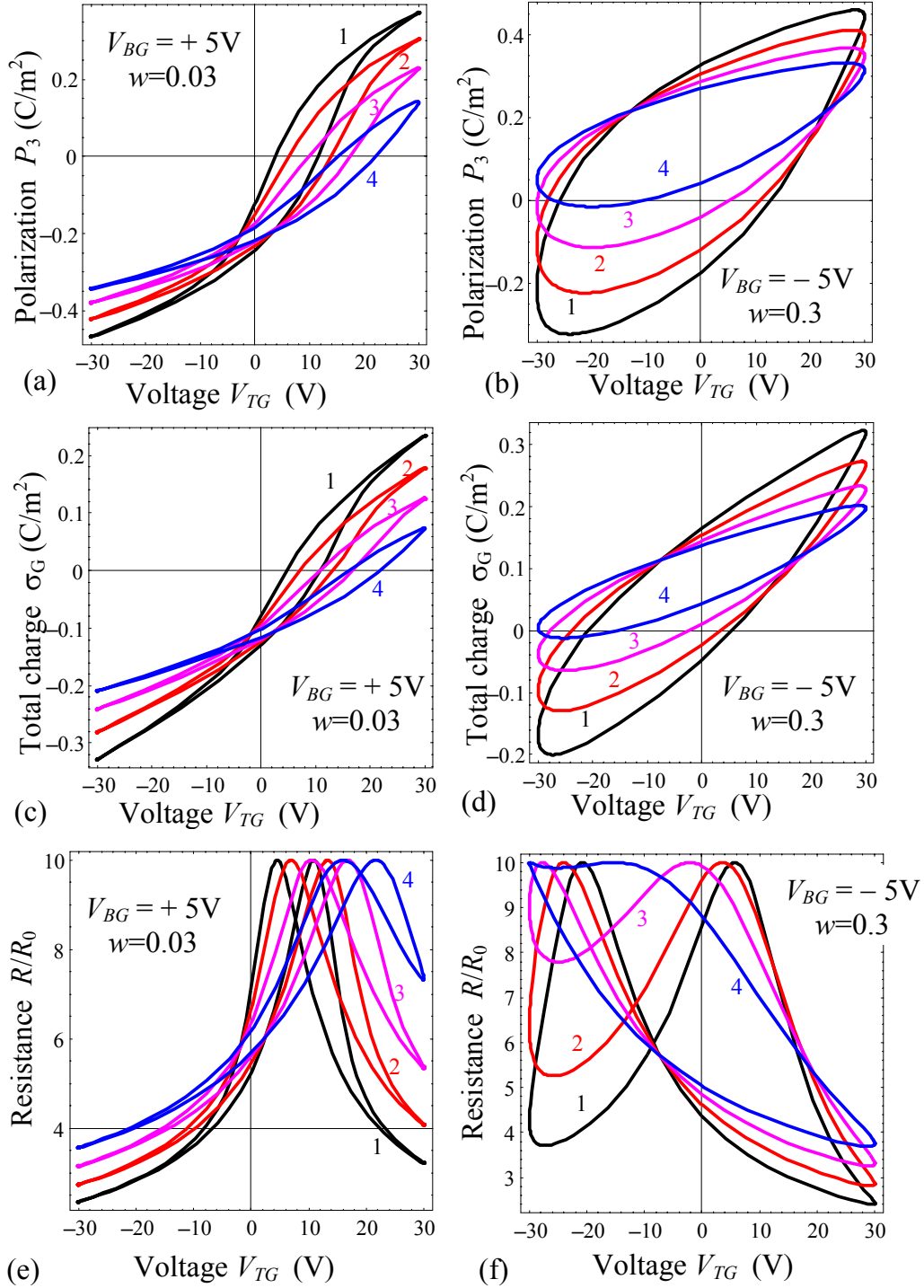

**Figure 4. Influence of multi-layer graphene thickness.** Hysteresis loops of LiNbO$_3$ ferroelectric polarization **(a, b)**, density of the total charge stored in multi-layer graphene **(c, d)** and its electro-resistance **(e, f)** calculated for different thickness of multi-layer graphene $d$=3, 4, 5, 6 nm (curves 1, 2, 3, 4). The frequency $w$=0.03 and back gate voltage $V_{BG}$= + 5 V for plots **(a)**, **(c)** and **(e)** (left column); $w$=0.3 and back gate voltage $V_{BG}$= − 5 V for plots **(b)**, **(d)** and **(f)** (right column). Sapphire dielectric thickness $h$=10 nm, screening radius $R_t$=3 nm, $n_{res}$ = 5×10$^{17}$ m$^{-2}$, ferroelectric film thickness $l$=20 nm.

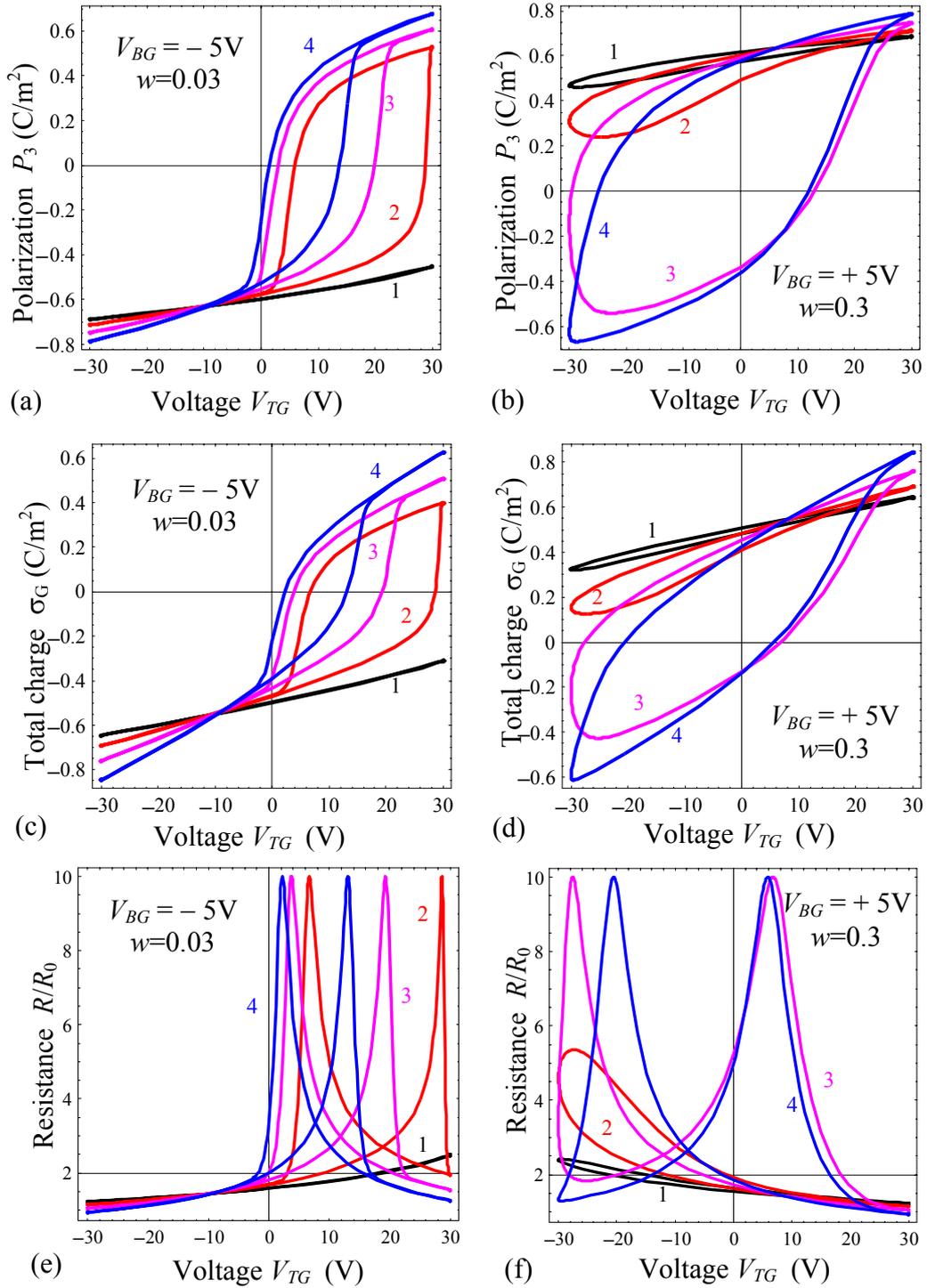

**Figure 5. Influence of screening radius value.** Hysteresis loops of LiNbO$_3$ ferroelectric polarization **(a, b)**, density of the total charge stored in multi-layer graphene **(c, d)** and its electro-resistance **(e, f)** calculated for different graphene screening radius $R_t$ =2.5, 2, 1.5 and 1.2 nm (curves 1, 2, 3, 4). The frequency w=0.03 and back gate voltage $V_{BG}$= - 5 V for plots **(a)**, **(c)** and **(e)** (left column); w=0.3 and back gate voltage $V_{BG}$= + 5 V for plots **(b)**, **(d)** and **(f)** (right column). Multi-layer graphene thickness d=3 nm, dielectric layer thickness h=1 nm, $n_{res}$ = 5×10$^{17}$ m$^{-2}$, ferroelectric film thickness l=10 nm.

**Figure 6** illustrates the coercive voltages of graphene charge in dependence on the dielectric layer, graphene and ferroelectric film thicknesses $h$, $d$ and $l$ correspondingly calculated for different frequencies $w$, zero and positive back gate voltage $V_{BG}$. Positive and negative coercive voltages are equal to each other for the case $V_{BG}=0$; the asymmetry appeared for nonzero $V_{BG}$ as anticipated. Finite size effects are evident, namely the coercive voltage decreases with $h$ increase and almost linearly increases with $d$ increase or $l$ increase. The coercive voltage increase with $l$ increase is almost linear. Unexpectedly the voltage increase with $l$ increase for $V_{BG}=0$ is very weak and reveals strongly asymmetric super-linear increase with $d$ increase. The effect can be explained from the asymmetry of the heterostructure electronic properties.

**Figure 7** shows the dependence of differential electro-resistance, $\delta R = (R_{max} - R_{min})/R_{min}$, in dependence on the dielectric layer, multi-layer graphene and ferroelectric film thicknesses $h$, $d$ and $l$ correspondingly calculated for different frequencies $w$ and fixed back gate $V_{BG}=5$ V. The way of $\delta R$ determination for resistance loops of different shape was the same as proposed by Zheng et al [7]. Dependence of $\delta R$ on $h$ has a pronounced maximum at low frequencies $w \leq 0.1$, which height decreases, width increases and position shifts to higher $h$ values with $w$ increase. At fixed $h=10$ nm the dependence of $\delta R$ on $d$ is monotonically decreasing at low frequencies $w \leq 0.1$, a diffuse maxima appears at $w = 0.3$. At fixed $h=5$ nm the dependence of $\delta R$ on $l$ has a pronounced maximum at low frequencies $w \leq 0.1$, which height decreases, width increases and position shifts to smaller $l$ values with $w$ increase. Note, that revealed maximum and its properties size dependence can be of great importance for optimization of GFeFET performances, especially for the application in nonvolatile memory devices of new generation, where large $\delta R$ is needed.

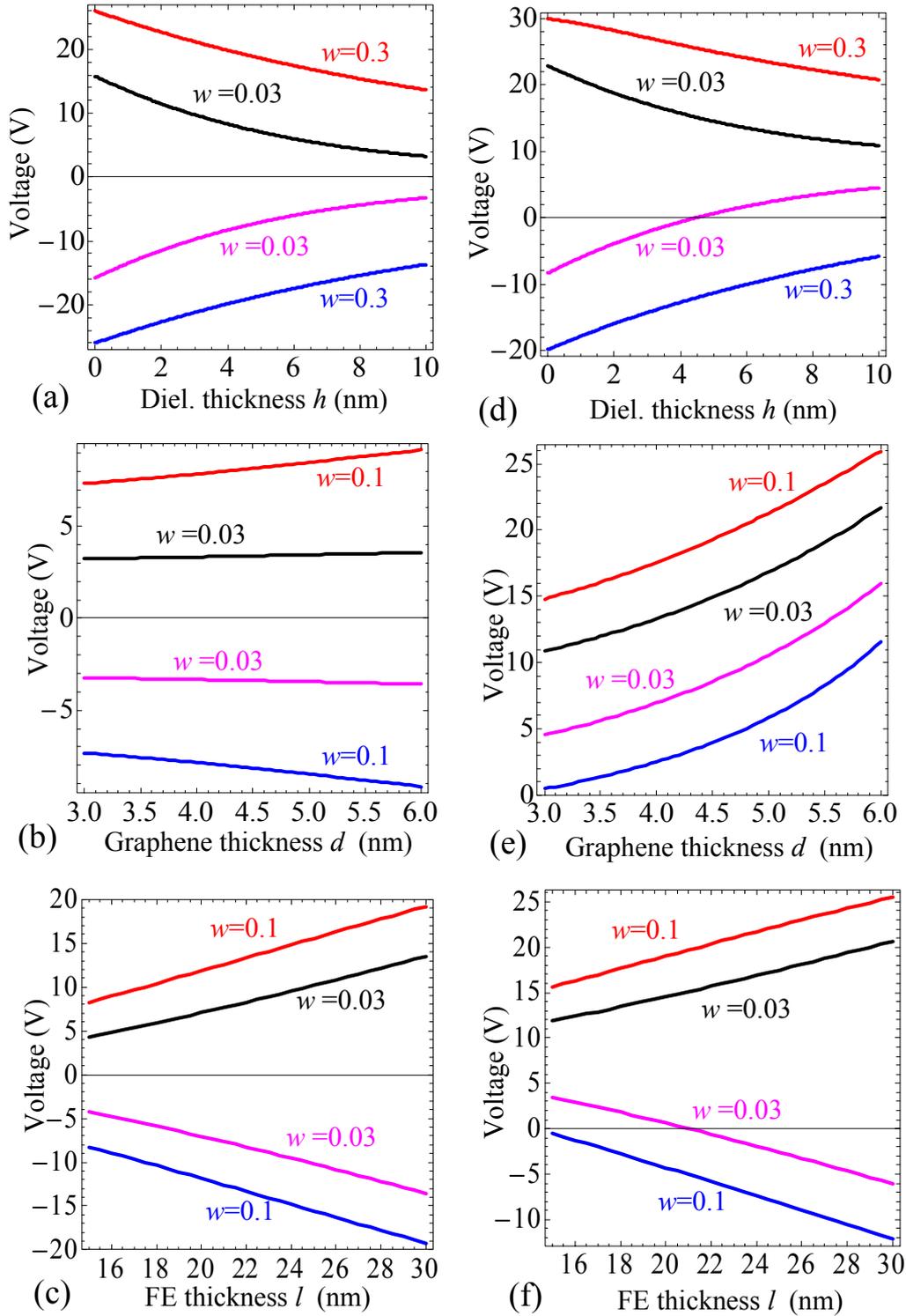

**Figure 6. Size effect of coercive voltages.** Graphene total charge left and right coercive voltages dependence on the dielectric layer, graphene and ferroelectric film thicknesses *h, d* and *l* correspondingly calculated for different frequencies *w* = 0.01 − 0.3 (labels near the curves). Plots (a), (b) and (c) are calculated for the zero back gate voltage, while (d), (e) and (f) with the back gate voltage 5V. For plots (a) and (d) *d* = 3 nm, *l* = 20 nm; (b) and (e) *h* = 10 nm, *l* = 20 nm; (c) and (f) *h* = 5 nm, *d* = 3 nm. Other parameters are the same as in the figure 4.

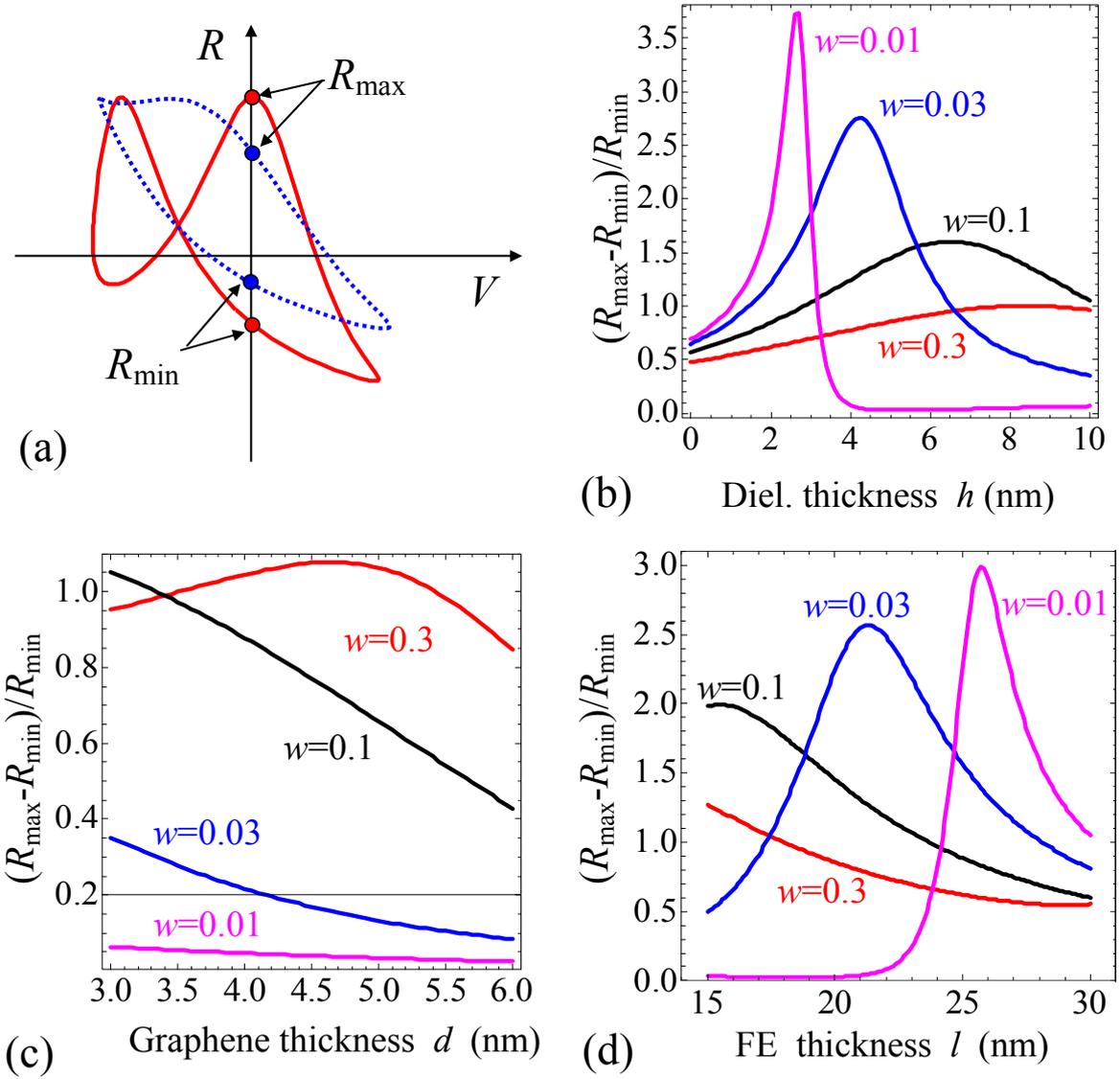

**Figure 7. Size effect on graphene electro-resistance. (a)** The way of $\delta R = (R_{max} - R_{min})/R_{min}$ determination for resistance loops of different shape. **(b-d)** Dependences of $\delta R$ on the dielectric layer, graphene and ferroelectric film thicknesses $h$, $d$ and $l$ calculated for different frequencies $w = 0.01 – 0.3$ (labels near the curves) are shown in the plots **(b)**, **(c)** and **(d)** correspondingly. Back gate voltage is $V_{BG}$=5 V; top gate voltage amplitude is 30 V and. Parameters for plot (b): $R_t$=3nm, $d$=3nm, $l$=20nm; (c) $R_t$=3nm, $h$=10nm, $l$=20nm; (d) $R_t$=3nm, $h$=5nm, $d$=3nm. Other parameters are the same as in the figure 4.

### 5. Domain structure impact on the equilibrium space charge redistribution in MLG

Let us finally consider a situation when the stripe domain structure is present in ferroelectric and electric dragging field $E_S$ is applied in x-direction to the **source electrode** located at graphene surface $z = -d$. For the case Eq.(2a) should be modified as

$$\varphi_G(x, y, z = -d) = V_{TG} - xE_S. \qquad (11)$$

Stationary solution for electric potential in a multi-domain state of ferroelectric is listed in **Appendix A.**

Below we study the space charge redistribution caused by domain stripes for the system "multi-layered graphene/dielectric/ferroelectric thin film" with the following parameters: Thomas-Fermi screening radius in graphene is $R_t$=0.5 nm, multi-layered graphene permittivity $\varepsilon_G$=15 (graphite), MLG thickness $d$=3.4 nm (N=10 graphene layers), sapphire layer thickness $h$=(5 – 50) nm, and compare with the case of intimate contact ($h$=0). Typical period of domain structure is $a$= (50 – 500) nm. Below we mainly use the value of 500 nm, that is much higher then graphene and dielectric thickness. Complex size effects possibly appeared in the case $a \sim h$ will be considered elsewhere. Applied fields $E_S = (0 - 10^3)$V/m are typical for the graphene-on-ferroelectric devices [20]. Full set of parameters are listed in the **Table 1**.

**Figures 8a-c** illustrate the influence of applied field $E_S$ and dielectric gap thickness on the space charge density distribution in the MLG modulated by ferroelectric domain stripes. Within Tomas-Fermi or Debye approximations the space charge density is proportional to the electric potential per Eq.(7), so their spatial modulations are in the antiphase. In the case of intimate contact ($h$=0) and ultra-thin dielectric layer (h<10 nm) the potential and space charge distributions are strongly affected by the depolarization field created by domain stripes. In the case $h$>10 nm and 0<$E_S$<50 V/m the charge density x-profile is quasi-harmonic function with maximums in the centre of domains and zero values on domain wall. The effect is manifested as quasi-rectangular modulation for $E_S$=0 and slightly incline steps for 0<$E_S$<500 V/m. The steps becomes smoother and smaller with $h$ increase and almost disappear at $h$=50 nm and $E_S$>50 V/m indicating that for the case external field effect appeared stronger that the depolarization field influence near the graphene upper surface $z = -d$. The event can be interpreted as a field-induced phase transition of the second order. The transition is local, because it is z-dependent. Actually depolarizing effect dominates with z-increase from graphene surface towards the dielectric layer. The characteristic scale of the effect disappearance is $R_t$ distance, which seems to be quite natural. However it seems essential, that for $E_S$ smaller then some critical value (< 1 kV/m) there are $p$ and $n$ domains in MLG with electrons and holes conductivity correspondingly, and the dominant scattering mechanism thus can be on $p$-$n$-junction potentials, corresponding the domain walls (see [23] and refs therein). For the higher fields there are no $p$ and $n$ domains no longer, which means the change of the dominant scattering mechanism (e.g. towards scattering on the charged impurities, present in the dielectric layer [2] or on the interface, which is generally less intensive than the long-range disorder scattering, described in [22]) and the dramatic increase of the MLG channel conductivity. The effect can be used in ferroelectric and sensoric applications.

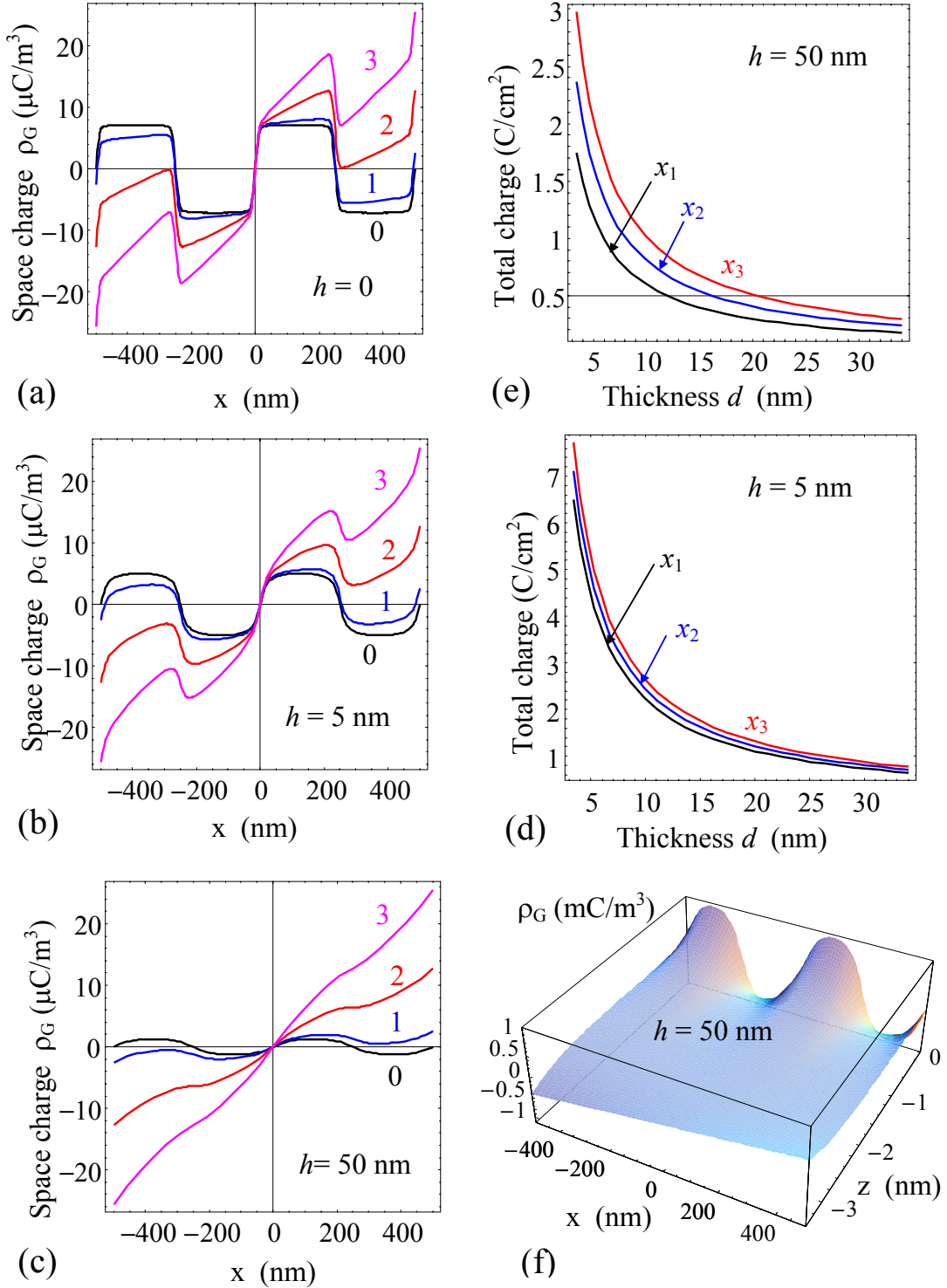

**Figure 8. Effect of domain stripes.** (a, b, c) Space charge x-distribution at $z = -d - R_t/2 = -3.15$ nm caused in MLG by ferroelectric domain stripes in Roshelle salt, with the presence of sapphire layer of different thickness $h$=0 (intimate contact), 5 nm and 50 nm on ferroelectric surface. Driving electric

field $E_S$=0, 10, 50 and 100V/m (curves with labels 0, 1, 2 and 3 correspondingly), $V_{TG}=V_{BG}$=0. Ferroelectric film thickness $l$=300 nm, domain stripe period $a$=500 nm. **(c-d)** Total charge dependence on MLG thickness $d$ calculated at different x-coordinates $x_1$= 125 nm, $x_2$= 725 nm and $x_3$= 1125 nm. Result is field-dependent, here we put $E_S$=1 kV/m. **(f)** 3D-map of the space charge calculated at $E_S$=1 kV/m.

The total charge density calculated at $E_S$=1 kV/m is shown in the **Figures 8d-e** for $h$=0 and dielectric layer thickness $h$=50 nm. Plots illustrate the monotonic decrease of the maximal total charge with graphene thickness increase, i.e. finite size effect. Saturation appears with $d$ increase. As one can see from the **Figure 8f** the pronounced charge modulation appeared near the top gate electrode; it is slightly affected by external field $E_S$.

## 7. Conclusion

The main physical factor responsible for the resisting memory effect in the considered heterostructure "multi-layer graphene–dielectric layer–ferroelectric film" is the hysteresis of the internal depolarizing electric field, which appears in the dielectric layer between multi-layer graphene and ferroelectric film. Depolarizing fields originated from the incomplete screening of ferroelectric film spontaneous polarization in the dielectric layer. Since the field reverses its sign under the spontaneous polarization reversal taking place when external field exceeds the coercive one (ferroelectric hysteresis), the process causes nonlinear hysteretic dynamics of graphene charge and electro-resistance. Main results obtained in the work are the following:

Analytical expressions for the acting electric field, graphene charge and electro-resistance are derived in the case when a uniformly polarized ferroelectric film is a perfect insulator, and its surface is free of screening charges. As anticipated the field value is defined by the by the thicknesses of dielectric layer $h$, multi-layer graphene film $d$, and ferroelectric film $l$ (in fact we report about pronounced size effects).

Versatile shape of graphene charge and electro-resistance hysteresis loops appears in the system depending on the layers thicknesses $h, l$ and $d$, and frequency and amplitude of electric voltage applied to the top gate (see **Table 2**). Note that calculated resistance loops are very similar to the experimental ones obtained earlier by Zheng et al [7] in GFeFET and correlate with the mechanism of the "direct" hysteresis of graphene channel resistivity, caused by repolarization of the dipoles in ferroelectric substrate [22].

Approach proposed for a single-domain film can be further evolved for the description of the space charge accumulation in graphene-on-ferroelectric allowing for the ferroelectric domain structure. In particular, when the ferroelectric substrate has evolved stripe domain structure and a driving electric field $E_S$ is applied along the graphene channel, the domain stripes of different polarity can induce domains with $p$ and $n$ conductivity in multi-layer graphene strip for $E_S$ smaller then some critical value. Thus for the case the dominant carrier scattering mechanism can be on randomly distributed $p$-$n$-junction potentials, which position correlates with domain walls in ferroelectric film. For the higher fields there are no pronounced $p$ and $n$ domains, only smeared ripples remained, which indicates the change of the dominant scattering mechanism and the dramatic increase of the graphene channel conductivity. The intriguing effect can open new possibilities for graphene-based sensors and explore the physical mechanisms underlying the operation of graphene field effect transistor with ferroelectric gating.

**Table 2. Size effects**

| Size | Influence on polarization $P_3$ | Influence on total charge $\sigma_G$ | Influence on electro-resistance $R$ and its variation $\delta R = (R_{max} - R_{min})/R_{min}$ |
|---|---|---|---|
| Dielectric layer thickness $h$ | Remnant polarization decreases with increasing $h$ | The total charge maximal difference (amplitude) decreases with increasing $h$. | Double peaks of resistance dependence get closer and become less "sharp" with $h$ increase. $\delta R(h)$ has a pronounced maximum at low frequencies, which height decreases, width increases and position shifts to higher $h$ values with frequency increase |
| | Coercive voltages, as well as the maximal loop width, decrease with increasing $h$. The loops lose the squire-like and acquire slim-like shape with $h$ increase. | | |
| Multi-layer graphene thickness $d$ | Remnant polarization decreases with increasing $d$ | The total charge amplitude decreases with increasing $d$ | The loops asymmetry increases with $d$ increase and double shape loops transform into a single shape. |
| | Coercive voltages, as well as the maximal loop width, decrease with increasing $d$. For layers of graphite (more than 30 graphene layer) hysteresis loops are not revealed | | |
| Ferroelectric thickness $l$ | Remnant polarization values decrease with $l$ decrease. | The total charge amplitude decreases with decreasing $l$. | The difference between two peaks of resistance increases with increasing $l$. $\delta R(l)$ has a maximum at low frequencies, which height decreases, width increases and position shifts to smaller $l$ values with frequency increase |
| | Coercive voltage values decrease with $l$ decrease. | | |
| Screening radius $R_t$ | Remnant polarization decreases with increasing $R_t$. | Total charge value becomes smaller with $R_t$ increase. | Resistance loops double shape transforms into a single one with $R_t$ increase. |
| | Coercive voltage values decreases with increasing $R_t$, hysteresis loop almost disappears when multi-layer graphene thickness $d$ becomes more than $2R_t$. Loops become smoother with increasing $R_t$. | | |

**Acknowledgments.** A.N.M., O.V.V. and E.A.E. acknowledge National Academy of Sciences of Ukraine, grant 35-02-14, and Center for Nanophase Materials Sciences, user projects CNMS 2013-293, CNMS 2014-270. Research for SVK was supported by the US Department of Energy, Basic Energy Sciences, Materials Sciences and Engineering Division.

**Appendix A**

**Stationary solution for electric potential in a multi-domain state of ferroelectric**

Let us finally consider a situation when the stripe domain structure is present in ferroelectric and electric dragging field $E_S$ is applied in x-direction to the *source electrode* located at graphene surface $z = -d$. For the case Eq.(2a) should be modified as $\varphi_G(x, y, z = -d) = V_{TG} - xE_S$.

Let us look for the general solution of Eqs.(1) with boundary conditions (2) using the Fourier transformation for transverse coordinate x, namely $\varphi_j(x,z) = \int_{-\infty}^{\infty} dk \varphi_j(k,z) e^{ikx}$, where the subscript $j = G, d, f$ means graphene, dielectric and ferroelectric correspondingly. The Fourier images of the potentials are:

$$\varphi_G(k,z) = \varphi_G^{TG} + \varphi_G^{BG} + \tilde{\varphi}_G^+(k)e^{qz} + \tilde{\varphi}_G^-(k)e^{-qz}, \quad (A.1a)$$

$$\varphi_d(k,z) = \varphi_d^{TG} + \varphi_d^{BG} + \tilde{\varphi}_d^+(k)e^{kz} + \tilde{\varphi}_d^-(k)e^{-kz}, \quad (A.1b)$$

$$\varphi_f(k,z) = \varphi_f^{TG} + \varphi_f^{BG} + \tilde{\varphi}_f(k)\sinh\left((L-z)\frac{k}{\gamma}\right). \quad (A.1c)$$

Here the contribution of the top and bottom gate voltages into the solution (1) is given by expressions:

$$\varphi_G^{TG} = V_{TG} e^{(d+z)/R_t} \frac{\varepsilon_d \varepsilon_G l(e^{-2z/R_t} + 1) + \varepsilon_{33}^f(\varepsilon_G h(e^{-2z/R_t} + 1) + \varepsilon_d R_t(e^{-2z/R_t} - 1))}{\varepsilon_d \varepsilon_G l(e^{2d/R_t} + 1) + \varepsilon_{33}^f(\varepsilon_G h(e^{2d/R_t} + 1) + \varepsilon_d R_t(e^{2d/R_t} - 1))}, \quad (A.2a)$$

$$\varphi_d^{TG} = \frac{2V_{TG} e^{d/R_t}(\varepsilon_d l + \varepsilon_{33}^f(h-z))\varepsilon_G}{\varepsilon_d \varepsilon_G l(e^{2d/R_t} + 1) + \varepsilon_{33}^f(\varepsilon_G h(e^{2d/R_t} + 1) + \varepsilon_d R_t(e^{2d/R_t} - 1))}, \quad (A.2b)$$

$$\varphi_f^{TG} = \frac{2V_{TG} e^{d/R_t}(l + h - z)\varepsilon_d \varepsilon_G}{\varepsilon_d \varepsilon_G l(e^{2d/R_t} + 1) + \varepsilon_{33}^f(\varepsilon_G h(e^{2d/R_t} + 1) + \varepsilon_d R_t(e^{2d/R_t} - 1))}, \quad (A.2c)$$

$$\varphi_G^{BG} = \frac{\varepsilon_{33}^f \varepsilon_d R_t V_{BG}(e^{(2d+z)/R_t} - e^{-z/R_t})}{\varepsilon_d \varepsilon_G l(e^{2d/R_t} + 1) + \varepsilon_{33}^f(\varepsilon_G h(e^{2d/R_t} + 1) + \varepsilon_d R_t(e^{2d/R_t} - 1))}, \quad (A.3a)$$

$$\varphi_d^{BG} = \frac{\varepsilon_{33}^f V_{BG}(\varepsilon_d R_t(e^{2d/R_t} - 1) + \varepsilon_G z(e^{2d/R_t} + 1))}{\varepsilon_d \varepsilon_G l(e^{2d/R_t} + 1) + \varepsilon_{33}^f(\varepsilon_G h(e^{2d/R_t} + 1) + \varepsilon_d R_t(e^{2d/R_t} - 1))}, \quad (A.3b)$$

$$\varphi_f^{BG} = V_{BG} - \frac{\varepsilon_d \varepsilon_G V_{BG}(e^{2d/R_t} + 1)(h + l - z)}{\varepsilon_d \varepsilon_G l(e^{2d/R_t} + 1) + \varepsilon_{33}^f(\varepsilon_G h(e^{2d/R_t} + 1) + \varepsilon_d R_t(e^{2d/R_t} - 1))}. \quad (A.3c)$$

The functions $\tilde{\varphi}_G^\pm(k) = \psi_G^\pm(k) + \sum_{m=0}^{\infty} \phi_G^\pm(k_m)\delta(k - k_m)$, $\tilde{\varphi}_d^\pm(k) = \psi_d^\pm(k) + \sum_{m=0}^{\infty} \phi_d^\pm(k_m)\delta(k - k_m)$ and $\tilde{\varphi}_f(k) = \psi_f(k) + \sum_{m=0}^{\infty} \phi_f(k_m)\delta(k - k_m)$. The division on $\psi_j^\pm(k)$ and $\phi_j^\pm(k_m)$ follows from the mathematical convenience to split the solution of the boundary problem with inhomogeneous boundary conditions in two parts, for the first part the "source" is proportional to ferroelectric polarization, while for the second part it is proportional to the applied field $E_S$. Therefore, components $\psi_j^\pm(k)$ are proportional to the source electric field amplitude $E_S$. Components $\phi_j^\pm(k_m)$ originate from the depolarization effect and are proportional to the polarization Fourier components.

Frequency spectrums are:

$$\phi_G^-(k) = \frac{2\varepsilon_d P_m \gamma \sinh(kl/\gamma)}{\varepsilon_0 Det(k)} e^{kh}, \quad \phi_G^+(k) = -\phi_G^-(k) e^{2qd}, \tag{A.4a}$$

$$\phi_d^-(k,\varepsilon_G) = \frac{-P_m \gamma \sinh(kl/\gamma)}{\varepsilon_0 k Det(k)} e^{kh}\left(\varepsilon_d k(e^{2qd}-1) - \varepsilon_G q(e^{2qd}+1)\right), \quad \phi_d^+(k,\varepsilon_G) = \phi_d^-(k,-\varepsilon_G) \tag{A.4b}$$

$$\phi_f(k) = \frac{-P_m\left(\varepsilon_d k(e^{2qd}-1)(e^{2kh}+1) + \varepsilon_G \gamma q(e^{2qd}+1)(e^{2kh}-1)\right)}{\varepsilon_0 k Det(k)} \tag{A.4c}$$

$$\psi_G^+(k,\varepsilon_G) = \frac{k e^{-k^2/4\alpha}}{2\sqrt{2}\alpha^{3/2}} \frac{iE_S e^{qd}}{Det(k)} \begin{pmatrix} \varepsilon_{33}^f\left(\varepsilon_d k(e^{2kh}+1) - \varepsilon_G q(e^{2kh}-1)\right)\cosh(kl/\gamma) + \\ \varepsilon_d\left(\varepsilon_d k(e^{2kh}-1) - \varepsilon_G q(e^{2kh}+1)\right)\gamma \sinh(kl/\gamma) \end{pmatrix}, \tag{A.4d}$$

$$\psi_G^-(k,\varepsilon_G) = -\psi_G^+(k,-\varepsilon_G), \tag{A.4e}$$

$$\psi_d^+(k,\varepsilon_d) = \frac{k e^{-k^2/4\alpha}}{\sqrt{2}\alpha^{3/2}} \frac{iE_S q e^{qd}}{Det(k)}\left(\varepsilon_{33}^f \cosh\left(\frac{kl}{\gamma}\right) - \varepsilon_d \gamma \sinh\left(\frac{kl}{\gamma}\right)\right), \quad \psi_d^-(k,\varepsilon_d) = -\psi_d^+(k,-\varepsilon_d), \tag{A.4f}$$

$$\psi_f(k) = \frac{k e^{-k^2/4\alpha}}{\sqrt{2}\alpha^{3/2}} \frac{-iE_S e^{qd+kh}}{Det(k)} q \varepsilon_d \varepsilon_G \gamma, \tag{A.4g}$$

$$Det(k) = \begin{pmatrix} \varepsilon_{33}^f \cosh\left(\frac{kl}{\gamma}\right)\left(\varepsilon_d k(e^{2kh}+1)(e^{2qd}-1) + \varepsilon_G q(e^{2kh}-1)(e^{2qd}+1)\right) \\ + \varepsilon_d \gamma \sinh\left(\frac{kl}{\gamma}\right)\left(\varepsilon_d k(e^{2kh}-1)(e^{2qd}-1) + \varepsilon_G q(e^{2kh}+1)(e^{2qd}+1)\right) \end{pmatrix}. \tag{A.4h}$$

For the case of periodic domain stripes with a period $a$, we can represent spontaneous polarization of ferroelectric in the form: $P_3(x) \approx \sum_{m=0}^{\infty} P_m \sin(k_m x)$, where $P_m \approx 4P_S/\pi(2m+1)$ and $k_m = (2m+1)(2\pi/a)$. Results for the case of a single-domain ferroelectric, can be obtained in the limit $k \to 0$. The wave number $q(k) = \sqrt{k^2 + R_t^{-2}}$. Later for numerical calculations convergence we formally substitute $xE_S \to e^{-\alpha x^2} xE_S$ with $\alpha=10^{10}$, that gives $e^{-\alpha x^2} \approx 1$ for $|x| < 1$ μm (the realistic value of graphene channel lengths in graphene-based FET devices is of μm-length order). Using the Delta-function representation, after elementary transformations we obtain the limit $\lim_{\alpha \to 0}\left(\frac{\exp(-k^2/4\alpha)k}{2\sqrt{2}\alpha^{3/2}}\right) = -\sqrt{\frac{\pi}{2}}\frac{d\delta(k)}{dk}$.